\documentclass[notitlepage]{article}
\usepackage{sw20lart}

\input{tcilatex}
\begin{document}

\title{Consistency, Amplitudes and Probabilities in Quantum Theory}
\author{Ariel Caticha \\
{\small Department of Physics, University at Albany-SUNY, Albany, NY 12222}}
\date{}
\maketitle

\begin{abstract}
Quantum theory is formulated as the only consistent way to manipulate
probability amplitudes. The crucial ingredient is a consistency constraint:
if there are two different ways to compute an amplitude the two answers must
agree. This constraint is expressed in the form of functional equations the
solution of which leads to the usual sum and product rules for amplitudes. A
consequence is that the Schr\"{o}dinger equation must be linear: non-linear
variants of quantum mechanics are inconsistent. The physical interpretation
of the theory is given in terms of a single natural rule. This rule, which
does not itself involve probabilities, is used to obtain a proof of Born's
statistical postulate. Thus, consistency leads to indeterminism.\newline
\end{abstract}

\section{Introduction}

In 1946 R. T. Cox gave an argument showing that once degrees of probability
are represented by real numbers there is a unique set of rules for inductive
reasoning, that is, for reasoning under conditions of insufficient
information.$^{\text{\cite{cox46}}}$ The crux of the argument is a
consistency requirement: if a probability can be computed in two different
ways, the two answers must agree. Cox expressed this consistency requirement
in the form of functional equations, the solution of which showed that the
rules for inductive reasoning coincide with the well-known rules of
probability theory. The importance of this achievement is twofold: First, it
legitimized viewing probability theory as an extended form of logic, a point
of view that goes back to Bernoulli and Laplace, was arguably held by Gibbs,
and which, more recently, has been forcefully advocated by Jaynes.$^{\text{%
\cite{jaynes83}}}$ Second, Cox's argument provides an explanation for the
uniqueness of probability theory, for its inevitability; any modifications
of the rules of probability theory will necessarily lead to inconsistencies,
and therefore be unsatisfactory.

This latter feature, the robustness of probability theory, is also shared by
quantum theory. The quest to explain the strange behavior of quantum systems
has, since the beginning, led to all sorts of attempts to modify the theory.
Two of the most central attributes of quantum theory, indeterminism and
linearity, have been the target of many such unsuccessful attempts. There
has been considerable progress on issues related to the possibility of
hidden variables and to the nature of statistical correlations.$^{\text{\cite
{bell87}}}$ Linearity violations have been considered as a means to resolve
the difficulties associated with the quantum mechanics of macroscopic
objects.$^{\text{\cite{debroglie50}}}$ Some authors have been motivated by
the fact that many linear physical theories are mere approximations to more
fundamental non-linear theories,$^{\text{\cite{bialynicki76}}}$ while others
were led either by the desire to test quantum mechanics ever more
stringently,$^{\text{\cite{weinberg89}}}$ or just to explore the curious
implications of non-linearity.$^{\text{\cite{gisin90}}}$ Such extensive
theoretical investigations have prompted several increasingly precise
experimental tests$^{\text{\cite{aspect82},\cite{shull80}}}$ which have
confirmed, at least for the time being, the robustness of quantum mechanics.

In this work we propose an approach to quantum theory using ideas inspired
by Cox's, although in a very different context. The result is the standard
quantum theory.$^{\text{\cite{dirac58},\cite{feynman48}}}$ The crux of our
argument is also a consistency requirement: if a probability amplitude can
be computed in two different ways, the two answers must agree. This
requirement is expressed in the form of functional equations, the solution
of which leads to the usual sum and product rules for quantum probability
amplitudes. In other words, quantum theory emerges as the unique way to
manipulate probability amplitudes consistently.

Next we obtain two important consequences. The first is that the equation
for time evolution, the Schr\"odinger equation, is necessarily linear. The
implication is that the question of whether non-linear versions of quantum
mechanics are at all possible should not be posed at the dynamical level of
the Schr\"odinger equation but rather at a much deeper kinematical level
requiring a reexamination of the use and utility of the concept of amplitude.

The second result addresses the issue of how does the knowledge of the
numerical value of an amplitude assist us in predicting the outcomes of
experiments. This question of the physical interpretation of an otherwise
abstract formalism is handled by proposing a very natural general rule which
applies to situations in which the result of an experiment is predicted with
certainty. Using this rule, which involves no probabilities, we obtain a
proof of Born's statistical postulate. The implication here is that a
quantum theory formulated in terms of consistently assigned amplitudes must
be indeterministic.

These two results, the proof of linearity and of Born's probability
interpretation, are not new. They have been anticipated within various
axiomatic approaches to quantum mechanics.$^{\text{\cite{jauch68}-\cite
{finkelstein63}}}$ For example, the fact that Born's postulate is actually a
theorem was independently discovered long ago by Gleason,$^{\text{\cite
{gleason57}}}$ by Finkelstein,$^{\text{\cite{finkelstein63}}}$ by Hartle$^{%
\text{\cite{hartle68}}}$ and by Graham.$^{\text{\cite{graham73}}}$ What is
new here is the manner in which the results are obtained; the new element is
the emphasis on consistency in a formulation where amplitudes, rather than
states or observables, are the central concept. From the pedagogical point
of view there is an advantage in sharing the intuitive appeal of Feynman's
path integrals.$^{\text{\cite{feynman48}}}$ Axiomatic methods, on the other
hand, tend to be considerably more abstract and mathematically
sophisticated; this is not in itself a defect but it does hinder their
accessibility.

To limit the risk that too general a treatment might obscure the simplicity
of the main ideas we will focus our attention on a simple example: a
particle with no spin or other internal structure; its only attribute is its
position. Furthermore, to avoid distractions with mathematical
technicalities (which might, in other contexts, be very relevant) we will
restrict the positions of the particle to sites on a discrete lattice. We
emphasize that these simplifications are not necessary. The generalization
to other systems involving more complicated configuration spaces is
straightforward. If one wants to describe a particle moving in a continuum
the modifications are rather trivial, a mere replacement of sums by
integrals; the case of a quantum field theory might not be as easy, but in
principle it should be doable as well.

In section 2 we consider various idealized experimental setups which will
test whether a particle moves from an initial starting point to a final
destination point. The use of these setups defines what statements or
propositions about the particle we are allowed to make. No mention of
observables beyond position is ever made, but there is a possibility of
combining simple setups into more complex ones. This is described by
introducing two operations, which we call $and$ and $or$, that allow us to
construct complex setups (or propositions) from simpler ones.

At first sight this approach to quantum theory might resemble other
axiomatic approaches. For example, in the quantum logic approach$^{\text{%
\cite{jauch68},\cite{hooker79},\cite{finkelstein63}}}$ propositions are also
defined operationally in terms of the setups that will test them. But there
are major differences, for example, an operation of central importance in
quantum logic is that of negating a proposition. In our approach negation is
never introduced. A comparison with the complex probability approach$^{\text{%
\cite{youssef91}}}$ also shows a similarity which on further analysis is,
again, proved superficial. Unlike the latter theory, our $and$ and $or$
operations are not the usual Boolean ones, although they do enjoy a
sufficient measure of associativity and distributivity to justify their
names. In fact, the set of statements allowed here is much more restricted
than in either of the two approaches mentioned.

In section 3 we seek a quantitative representation of the $and$ and $or$
operations, i.e., a representation of the possible relations among various
experimental setups. This is done by assigning a complex number to each
setup in such a way that relations among setups translate into relations
among the corresponding complex numbers. The assignment is highly
constrained by a consistency requirement, expressed in terms of functional
equations, that if the complex number associated to a given setup can be
computed in more than one way, the various answers must agree. Solving the
consistency constraints shows that all representations of $and$ and $or$ are
actually equivalent to each other and that there is a particular choice that
is singled out by its convenience. With this choice the $and$ and $or$
operations are represented by the product and sum of complex numbers, i.e.,
the product and sum of probability amplitudes.$^{\text{\cite{caticha97}}}$

After introducing, in section 4, the concept of a state described by a wave
function, we show how the product and sum rules imply the linearity of the
Schr\"odinger equation. Then we address the issue of the physical
interpretation of the formalism, that is, of how probability amplitudes are
to be used in the prediction of outcomes of experiments and, in sections 5
and 6, we give a proof of Born's statistical postulate. Final comments
appear in section 7.

\section{What can we say about a simple particle?}

Suppose the only experiments we can perform are those that can detect the
presence or absence of the particle in a sufficiently small region of
space-time around an event $x=\left(\vec{x},t\right)$. Later, in section 6,
we will argue that this is not as restrictive as one might at first think.
The propositions in which we are interested will typically describe motion.
The simplest statement of this sort, ``the particle moves from $x_i$ to $x_f$%
,'' which we will denote by $[x_f,x_i]$, can be tested by preparing a
particle at $x_i$ and placing a detector at $x_f$. Our eventual goal is that
of predicting the likelihood of a positive outcome of a test of $[x_f,x_i]$.

To analyze this problem we consider placing various obstacles in the path of
the particle. This leads to propositions involving various constraints
intermediate between $x_i$ and $x_f$. Consider, for example, ``the particle
goes from $x_i$ to $x_f$ via the intermediate point $x_1$'' (we assume that $%
t_i<t_1<t_f$) or, equivalently, ``the particle goes from $x_i$ to $x_1$ and
from there to $x_f$''. This we will denote by $[x_f,x_1,x_i]$. How could we
test this proposition? We will certainly have to prepare the particle at the
starting point $x_i$ and place a detector at the final destination point $%
x_f $, but we cannot place a second detector at $x_1$; our particle is a
delicate microscopic object, and our detectors are clumsy macroscopic
devices that will totally alter the nature of the motion. All detections
should be kept to the bare minimum: just one detection at the final
destination point.

To carry out a test of $[x_f,x_1,x_i]$ we will imagine an experimental setup
with a source at $x_i$, a detector at the final destination $x_f$ and some
sort of device which implements the constraint at $x_1$. Needless to say, we
deal here with a highly idealized conceptual device used as an aid for
reasoning rather than for actual experimentation; such devices are not
unusual in theoretical physics. We imagine first an extended obstacle which
blocks all paths that the particle could have taken through some arbitrary
spacetime region (see fig. 1a). This already represents a considerable
idealization; a more realistic obstacle would have fuzzy edges, regions of
partial transparency rather than total opacity, and so on, but let us
nevertheless proceed. The complications due to the arbitrary shape can be
alleviated by imagining our obstacle as a succession of simpler obstacles
each operating at a single time (see fig. 1b). The next step in
idealization, shown in fig. 1c, is one of these single-time obstacles of
infinitesimal spatial extent: it blocks all the paths passing through the
spacetime point $x_1$.

\FRAME{ftFU}{5.0004in}{3in}{0pt}{\Qcb{\textbf{(a)} A generic obstacle of
irregular shape and duration is placed in the path of the particle from $x_i$
to $x_f$. \textbf{(b)} The obstacle can be envisioned as a succession of
idealized obstacles of irregular shapes but essentially no duration. \textbf{%
(c)} An obstacle of infinitesimal extent; it blocks all the paths passing
through a given spacetime point. }}{\Qlb{fig1}}{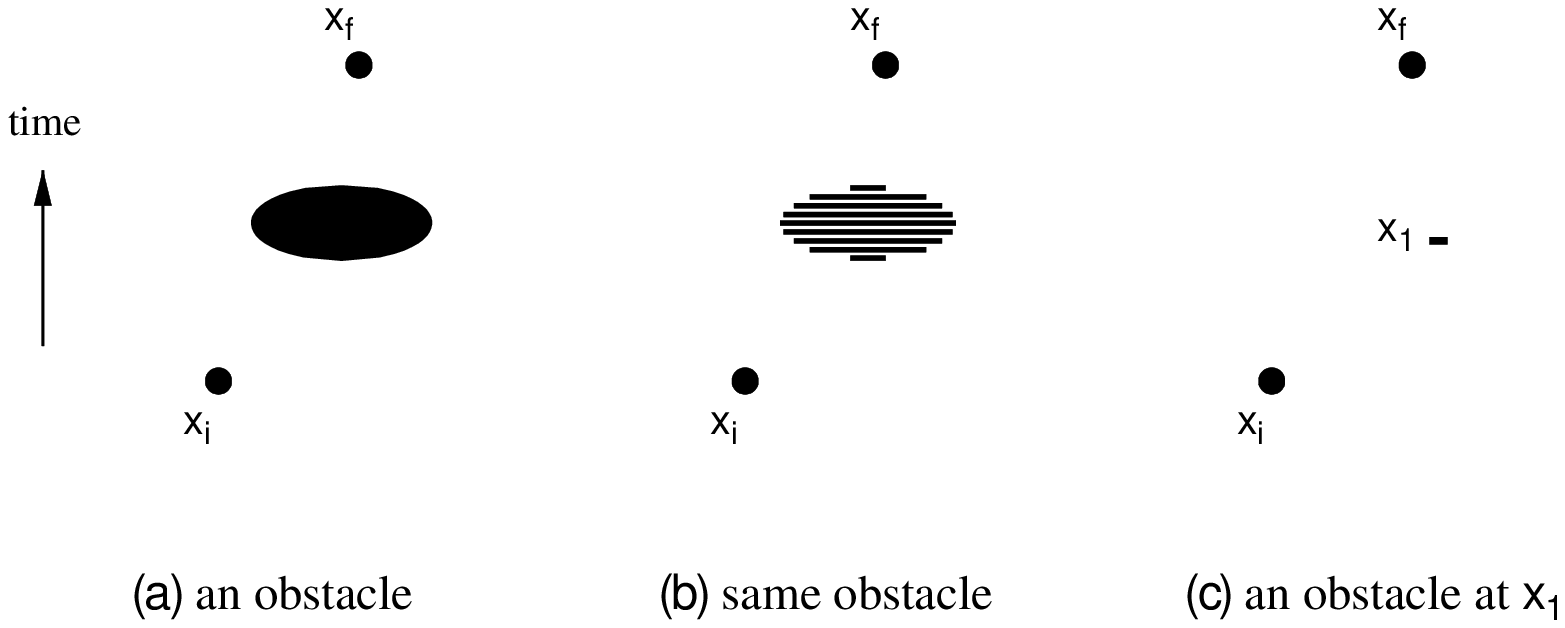}{\special{language
"Scientific Word";type "GRAPHIC";maintain-aspect-ratio TRUE;display
"USEDEF";valid_file "F";width 5.0004in;height 3in;depth 0pt;original-width
558.0625pt;original-height 430.625pt;cropleft "0";croptop "0.8882";cropright
"1";cropbottom "0.2248";filename
'C:/A-PAPERS/Q-Theory/Capqt/Fig1.eps';file-properties "XNPEU";}}

In fig. 2a and 2b we show the setups needed to test $[x_f,x_i]$ and $%
[x_f,x_1,x_i]$. The obstacle that implements the constraint at $x_1$ is the
complement of the infinitesimal obstacle of fig. 1c. This idealized device
we will call a ``filter.'' It suddenly appears at time $t_1$, blocking the
particle everywhere in space except for a small ``hole'' around the point $%
\vec{x}_1$, through which the particle may pass undisturbed. The filter
lasts an infinitesimally short interval and then, just as suddenly, it
disappears. The net result is that the filter prevents any motion from $x_i$
to $x_f$ except via the intermediate point $x_1$. If we want to impose more
constraints, as in $[x_f,x_2,x_1,x_i]=$``the particle goes from $x_i$ to $%
x_f $ via the intermediate events $x_1$ and $x_2$,'' we introduce two
filters, one at time $t_1$ and another at $t_2$ with holes at $\vec{x}_1$
and $\vec{x}_2$ respectively. The use of an infinite number of filters would
allow us to specify completely the path followed by the particle.

\FRAME{ftFU}{5.0004in}{3in}{0pt}{\Qcb{Examples of simple propositions: 
\textbf{(a)} ``the particle moves from $x_i$ to $x_f$'', \textbf{(b)} ``the
particle moves from $x_i$ to $x_1$ and from there to $x_f$'', and \textbf{(c)%
} ``the particle goes from $x_i$ to $x_f$ via $x_1$ or $x_1^{\prime }$''.}}{%
\Qlb{fig2}}{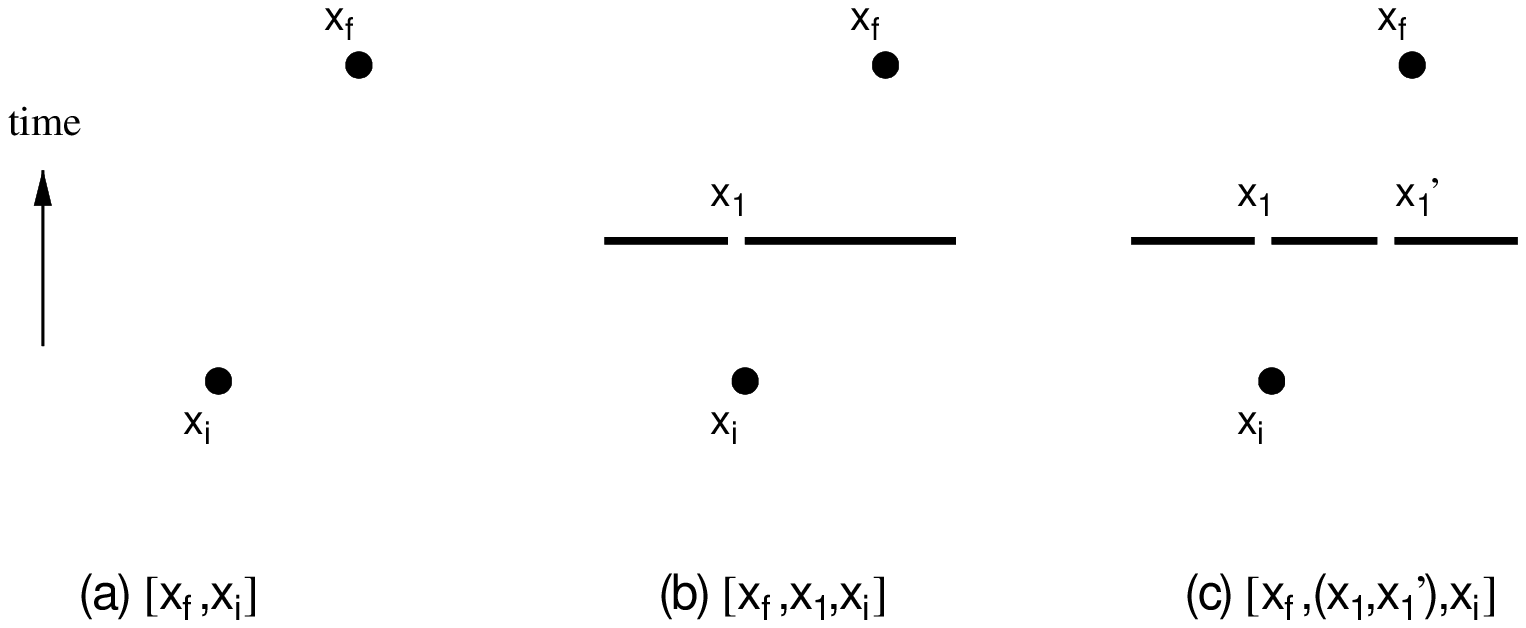}{\special{language "Scientific Word";type
"GRAPHIC";maintain-aspect-ratio TRUE;display "USEDEF";valid_file "F";width
5.0004in;height 3in;depth 0pt;original-width 558.0625pt;original-height
430.625pt;cropleft "0";croptop "0.8882";cropright "1";cropbottom
"0.2247";filename 'C:/A-PAPERS/Q-Theory/Capqt/Fig2.eps';file-properties
"XNPEU";}}

The conceptual device of using these idealized filters allows us to
introduce yet another kind of setup or proposition. Suppose that instead of
having one hole in the filter at $t_1$ we open two holes, one at $\vec{x}_1$
and another at $\vec{x}_1^{\prime}$(see fig. 2c). This physical situation is
one which one might classically describe as ``the particle goes from $x_i$
to $x_f$ via point $x_1$ or $x^{\prime}_1$.''$\,$ Such a proposition we will
denote by $[x_f,x_1,x^{\prime}_1,x_i]$. (We will generally use subscripts to
label the times at which events or filters occur and superscripts or primes
to distinguish events or holes which happen at the same time but at
different locations.) Although it is not quite necessary, for the sake of
clarity, we may wish to write $[x_f,(x_1,x^{\prime}_1),x_i]$ where we have
grouped together events which, being simultaneous, represent holes in the
same filter.

The propositions we will consider will all be of the general form 
\begin{equation}  \label{propdef}
a=[x_f,s_N,s_{N-1},\ldots,s_2,s_1,x_i]\text{ ,}
\end{equation}
where $s_n=(x_n,x^{\prime}_n,x^{\prime\prime}_n,\ldots)$ denotes a filter at
time $t_n$, intermediate between $t_i$ and $t_f$, with holes at $\vec{x}_n,%
\vec{x}^{\prime}_n,\vec{x}^{\prime\prime}_n,\ldots$ Statements such as $%
[s_N,s_{N-1},\ldots,s_2,s_1,x_i]$ or $[x_f,s_N,s_{N-1},\ldots,s_2,s_1]$ are
not allowed. Two propositions will be considered equal when they represent
the same experimental setup, i.e., the same distribution of filters and
holes.

Notice that equation (\ref{propdef}) incorporates two crucial features of
quantum theory: First, the allowed setups involve a single initial and a
single final event where we can place a source and a detector.$^{\text{\cite
{griffiths96}}}$ This is a recognition that measurements and other
interactions with macroscopic devices that induce uncontrollable
disturbances must be avoided. Second, there is a one to one correspondence
between the allowed statements and the idealized experimental setups with
which we could test those propositions: all propositions are testable. In
fact, we are identifying propositions with setups; this is a recognition
that no statements can be made about the particle by itself independently of
the experimental context.

From now on the words 'proposition' and 'setup' will be used
interchangeably. In fact, we prefer the latter and will use it more often
because, first, its use helps emphasize that the goal is to find out whether
the detector at $x_f$ will fire or not. Second, by avoiding statements about
the particle itself we hope to eliminate misconceptions about what the
particle is and what it is actually doing between source and detector. We
are not saying that the particle is either a point particle or a wave, or
both, or neither. We are not saying that it went through either one hole or
through another, or even that it went through both holes at the same time.
In fact, beyond the fact that the particle is capable of being emitted and
detected we are not assuming much at all.

In attempting to predict the result of tests it seems reasonable to assume
that if two propositions are related in some way (one proposition might, for
example, be testable using a part of the setup used for the other), then
information about one should be relevant to predictions about the other. Our
next step will be to exhibit relations of this sort. This will allow us to
use simple setups to build more complex ones, and conversely, also to
analyze complex setups into simpler ones.

The basic relations that we wish establish are of two kinds. The first kind
arises when two setups $a$ and $b$ can be placed in immediate succession.
This results in a third setup, obviously ``related'' to the first two, which
we will denote by $ab$. Notice that this operation, which we will call $and$%
, cannot be used to combine any two arbitrarily chosen setups $a$ and $b$.
It is only when the destination point of the earlier setup coincides with
the source point of the later setup that the combined $ab$ is an allowed
setup; $a$ and $b$ must be consecutive. The simplest instance of this is 
\begin{equation}
\lbrack x_f,x_1][x_1,x_i]=[x_f,x_1,x_i],
\end{equation}
\newline
and another example is shown in fig. 3a. In general, 
\begin{equation}
\lbrack x_f,s_N,\ldots ,s_{n+1},x_n][x_n,s_{n-1},\ldots
,s_1,x_i]=[x_f,s_N,\ldots ,x_n,\ldots ,s_1,x_i].  \label{and1}
\end{equation}
\newline
Conversely, any proposition with a filter containing a single hole can be
decomposed into two consecutive propositions. In the left member of equation
(\ref{and1}) it is important that all $t_1,\ldots ,t_{n-1}$ happen before $%
\,t_n$ and that $t_{n+1},\ldots ,t_N$ happen after $t_n$, otherwise the two
setups overlap and are not consecutive.

\FRAME{ftFU}{5.0004in}{3in}{0pt}{\Qcb{Two examples of using $and$ and $or$
to construct complex propositions out of simpler ones.}}{\Qlb{fig3}}{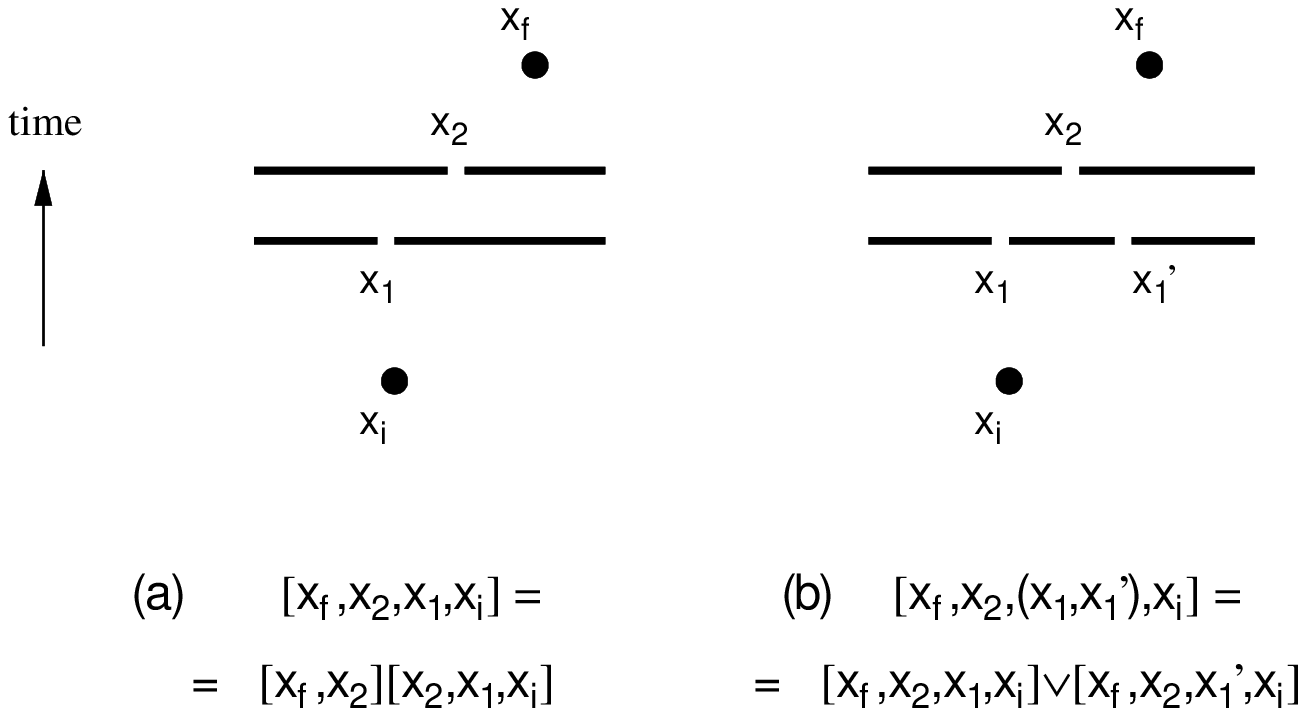%
}{\special{language "Scientific Word";type "GRAPHIC";maintain-aspect-ratio
TRUE;display "USEDEF";valid_file "F";width 5.0004in;height 3in;depth
0pt;original-width 558.0625pt;original-height 430.625pt;cropleft "0";croptop
"0.8907";cropright "1";cropbottom "0.2249";filename
'C:/A-PAPERS/Q-Theory/Capqt/Fig3.eps';file-properties "XNPEU";}}

\ The second useful kind of relation we consider arises when two setups $%
a^{\prime }$ and $a^{\prime \prime }$ are identical except on one single
filter where none of the holes of $a^{\prime }$ overlap any of the holes of $%
a^{\prime \prime }$. We may then form a third setup $a$, denoted by $%
a^{\prime }\vee a^{\prime \prime }$, which includes the holes of both $%
a^{\prime }$ and $a^{\prime \prime }$. A simple instance of this operation,
which we will call $or$, occurs in a ``two-slit'' experiment (see fig. 2c), 
\begin{equation}
\lbrack x_f,x_1^{\prime },x_i]\vee [x_f,x_1^{\prime \prime
},x_i]=[x_f,\left( x_1^{\prime },x_1^{\prime \prime }\right) ,x_i],
\end{equation}
\newline
and another example is shown in fig. 3b. The general case is 
\begin{equation}
\lbrack x_f,\ldots ,s_n^{\prime },\ldots ,x_i]\vee [x_f,\ldots ,s_n^{\prime
\prime },\ldots ,x_i]=[x_f,\ldots ,s_n,\ldots ,x_i],  \label{or1}
\end{equation}
where 
\[
s_n^{\prime }=(x_n^{\prime 1},x_n^{\prime 2},\ldots ,x_n^{\prime j}),\text{%
\quad }s_n^{\prime \prime }=(x_n^{\prime \prime 1},x_n^{\prime \prime
2},\ldots ,x_n^{\prime \prime k}), 
\]
and 
\[
s_n=(x_n^{\prime 1},\ldots ,x_n^{\prime j},x_n^{\prime \prime 1},\ldots
,x_n^{\prime \prime k}). 
\]
\newline
Again, notice that it is only for special choices of setups $a$ and $b$ that
this $or$ operation will result in an allowed setup $a\vee b$.

The two symbols we have introduced, $and$ and $or$, represent our presumed
ability to construct more complex setups out of simpler ones. If one
considers them as operations it is natural to ask if there are any rules
that should be followed to manipulate them consistently. To obtain these
rules we follow the principle, mentioned earlier, that two propositions are
equal when they represent the same setup of filters and holes. The first
rule is that the $or$ operation is commutative 
\begin{equation}
a\vee b=b\vee a.
\end{equation}
\newline
For the $and$ operation, however, there is an asymmetry implicit in the idea
of placing setups in succession, one setup is the earlier one. It is
convenient to incorporate this feature into the notation: if \ $ab$ is an
allowed setup, \ 
\begin{equation}
ab\neq ba,  \label{commor}
\end{equation}
\newline
because $ba$ is not allowed.

Next, we can see that both $and$ and $or$ enjoy a certain amount of
associativity. For example, given three consecutive setups $a$, $b$ and $c$
we can write 
\begin{equation}
\left( ab\right) c=a\left( bc\right) \equiv abc,  \label{assocand}
\end{equation}
\newline
provided $ab$ and $bc$ are allowed. In this case $\left( ab\right) c$ and $%
a\left( bc\right) $ are automatically allowed but $ac$ is not because $a$
and $c$ are not consecutive. Similarly, for the $or$ operation we have \ 
\begin{equation}
\left( a\vee b\right) \vee c=a\vee \left( b\vee c\right) \equiv a\vee b\vee
c,  \label{assocor}
\end{equation}
\newline
provided all four setups $\left( a\vee b\right) $, $\,\left( b\vee c\right) $%
, $\,\left( a\vee b\right) \vee c$ and $a\vee \left( b\vee c\right) $ are
allowed. In this case eq. (\ref{assocor}) is also equal to $\left( a\vee
c\right) \vee b$. Notice that any differences between the three setups $a$, $%
b$, and $c$ must be found in one and the same filter. Otherwise \ $\left(
a\vee b\right) \vee c\neq a\vee \left( b\vee c\right) $ because if the
member on the left is allowed the one on the right is not.

The last important rule is that of distributivity. This may take the form 
\newline
\begin{equation}
a(b\vee c)=\left( ab\right) \vee \left( ac\right) \quad \quad \text{or}\quad
\quad (b\vee c)a=\left( ba\right) \vee \left( ca\right) .  \label{distrib}
\end{equation}
\newline
Which equality holds, if any, will depend on whether the relevant
propositions are allowed. Both equalities cannot hold simultaneously.

An important illustration of the use and utility of the $and$ and the $or$
operations arises from the observation that a single filter that is totally
covered with holes is equivalent to having no filter at all. In other words,
the absence of a filter at time $t_1$ is a special kind of filter $\sigma _1$
which may be freely introduced into any proposition (provided $t_i<t_1<t_f$%
). For example, using eqs. (\ref{and1}) and (\ref{or1}), we have 
\begin{equation}
\lbrack x_f,x_i]=[x_f,\sigma _1,x_i]=\stackunder{\text{all}\,\vec{x}_1}{\vee 
}[x_f,x_1,x_i]=\stackunder{\text{all}\,\vec{x}_1}{\vee }([x_f,x_1][x_1,x_i]),
\label{sigmat}
\end{equation}
\ \newline
and, introducing additional $\sigma $ filters at times $t_2,\ldots ,$ $t_N$
we get 
\begin{equation}
\lbrack x_f,x_i]=\stackunder{\text{all}\,\vec{x}_N}{\vee }\cdots \stackunder{%
\text{all}\,\vec{x}_2}{\vee }\,\,\stackunder{\text{all}\,\vec{x}_1}{\vee }%
([x_f,x_N]\cdots [x_2,x_1][x_1,x_i]).
\end{equation}
\newline
This shows how motion over a long distance can be analyzed in terms of
motion over shorter steps.

One cannot fail to see some similarity between our quantum $and$ and $or$
operations with the Boolean operations and and or which also happen to be
commutative, associative and distributive and are also used to construct
more complex propositions out of simpler ones. But the similarity ends
there: The quantum $and$ and $or$ introduced here are not logical but rather
physical connectives, they are used to describe the relative dispositions of
various pieces of equipment. Also, and perhaps more important, is the fact
that the Boolean operations (and this applies as well to the and and or
introduced in quantum logics) can connect any two arbitrary propositions,
while the conditions on allowed setups impose severe restrictions on the
action of the quantum $and$ and $or$.

\section{Amplitudes: the sum and product rules}

Our goal is to predict the outcomes of experiments and the strategy is to
establish a network of relations among setups in the hope that information
about some setups might be helpful in making predictions about others. Our
next step will be to obtain a quantitative representation of these relations.

Suppose each setup $a$ is assigned a complex number $\phi (a)$. By a
`representation' we mean that the assignment of $\phi $s is such that
relations among physical setups should translate into relations among the
complex numbers associated to them. Why should such a representation exist?
It need not, but all of physics consists of representing elements of
reality, or relations among these elements, or our information about them,
by mathematical objects. The existence of such representations may be
mysterious, but it is not surprising; there are too many examples. A second,
simpler question is why do we seek a representation in terms of complex
numbers? Again, no answer here; this is an unexplained feature of quantum
theory. It seems that a single complex number is sufficient to convey the
physically relevant information about a setup.

To be specific consider a ``double-slit'' experiment. The relation between $%
[x_f,(x_1,x_1^{\prime }),x_i]$ and its components $[x_f,x_1,x_i]$ and $%
[x_f,x_1^{\prime },x_i]$ will be represented as a relation between the
complex numbers $\phi (x_f,(x_1,x_1^{\prime }),x_i)$, $\phi (x_f,x_1,x_i)$
and $\phi (x_f,x_1^{\prime },x_i)$ corresponding to them. What we require is
that there exist a function $S$ such that 
\begin{equation}
\phi \left( x_f,\left( x_1,x_1^{\prime }\right) ,x_i\right) =S\left( \phi
\left( x_f,x_1,x_i\right) ,\phi \left( x_f,x_1^{\prime },x_i\right) \right) ,
\end{equation}
and that this same function apply to any other setups that are similarly
related. More generally, if the setups associated to $a$ and to $a^{\prime }$
are such that $a\vee a^{\prime }$ is an allowed setup then 
\begin{equation}
\phi \left( a\vee a^{\prime }\right) =S\left( \phi \left( a\right) ,\phi
\left( a^{\prime }\right) \right) .
\end{equation}
Thus, the function $S$ is a representation of the relation $or$.

The requirement that $S$ should exist is a strong constraint on the allowed
assignment of $\phi $s. Consider for example the number $\phi $ assigned to $%
a\vee a^{\prime }\vee a^{\prime \prime }$. Using associativity this can be
calculated in two different ways, either as $\phi \left( \left( a\vee
a^{\prime }\right) \vee a^{\prime \prime }\right) $ or as $\phi \left( a\vee
\left( a^{\prime }\vee a^{\prime \prime }\right) \right) $. Consistency
requires that the two ways agree, 
\begin{equation}
S\left( \phi \left( a\vee a^{\prime }\right) ,\phi \left( a^{\prime \prime
}\right) \right) =S\left( \phi \left( a\right) ,\phi \left( a^{\prime }\vee
a^{\prime \prime }\right) \right) .
\end{equation}
\newline
Using $S$ once again one obtains the following consistency constraint 
\begin{equation}
S\left( S\left( u,v\right) ,w\right) =S\left( u,S\left( v,w\right) \right) ,
\label{constrs}
\end{equation}
\ where we have let $\phi \left( a\right) =u$, $\phi \left( a^{\prime
}\right) =v$, and $\phi \left( a^{\prime \prime }\right) =w$.

One can check, by substitution, that the associativity constraint, eq.(\ref
{constrs}), is satisfied if 
\begin{equation}
S\left( u,v\right) =\xi ^{-1}\left( \xi \left( u\right) +\xi \left( v\right)
\right) \quad \quad \text{or}\quad \quad \xi \left( S\left( u,v\right)
\right) =\xi \left( u\right) +\xi \left( v\right) ,  \label{xi1}
\end{equation}
where $\xi $ is an arbitrary function. In the appendix we give a proof
(similar to Cox's$^{\text{\cite{cox46}}}$) that this is also the general
solution. In other words, eq.(\ref{xi1}) tells us what forms the function $S$
may take, and conversely, that if the function $S$ exists then there must
also exist another function $\xi $, calculable from $S$, such that 
\begin{equation}
\xi \left( \phi \left( a\vee a^{\prime }\right) \right) =\xi \left( \phi
\left( a\right) \right) +\xi \left( \phi \left( a^{\prime }\right) \right) .
\end{equation}
This is remarkable. It immediately suggests that instead of the original
representation in terms of the complex numbers $\phi \left( a\right) $, we
should opt for an equivalent, simpler and more convenient representation in
terms of the numbers $\xi \left( \phi \left( a\right) \right) $. In other
words, the consistent assignment of complex numbers $\xi \left( a\right) $
to propositions $a$ can always be done so that the $or$ operation is
represented by a simple sum rule, \ 
\begin{equation}
\xi \left( a\vee a^{\prime }\right) =\xi \left( a\right) +\xi \left(
a^{\prime }\right) .  \label{sumxi}
\end{equation}
\newline
In this representation $S$ is addition.

Next we turn our attention to the representation of the $and$ operation.
From this point onward Cox's treatment and ours differ. Cox focused on the
operation of negating a proposition, and was thus led to consider the
consistency requirement ensuing from the possibility of double negation.
Negation is not an operation available to us, we rather choose to
concentrate on the associative and distributive properties of $and$.

Consider for example, a particle that goes from an initial $x_i$ to a final $%
x_f$ via an intermediate point $x$. We want to represent the relation
between $[x_f,x,x_i]$ and its components $[x_f,x]$ and $[x,x_i]$ as a
relation between the complex numbers $\xi \left( x_f,x,x_i\right) $, $\xi
\left( x_f,x\right) $ and $\xi \left( x,x_i\right) $. We then require that
there exist a function $P$ such that 
\begin{equation}
\xi \left( x_f,x,x_i\right) =P\left( \xi \left( x_f,x\right) ,\xi \left(
x,x_i\right) \right) ,
\end{equation}
and that the same function $P$ apply to any other propositions that are
similarly related. Specifically, if $ab$ is any allowed proposition we
require that \ 
\begin{equation}
\xi \left( ab\right) =P\left( \xi \left( a\right) ,\xi \left( b\right)
\right) ,
\end{equation}
so that the function $P$ is a representation of the $and$ operation.

The functional form of $P$ is highly constrained by the requirement that $P$
should exist for any $a$ and $b$ such that $ab$ is allowed. We can repeat
the argument we used earlier for the $or$ operation: the number $\xi$
associated to the (allowed) proposition $abc$ can be computed in two ways,
either as $\xi\left(\left(ab\right)c\right)$ or as $\xi\left(a\left(bc%
\right)\right)$, and these should agree. Therefore, $P$ must satisfy the
associativity constraint

\begin{equation}
P\left( P\left( u,v\right) ,w\right) =P\left( u,P\left( v,w\right) \right) .
\label{constrp}
\end{equation}
\newline
Furthermore, $and$ and $or$ are not unrelated: using distributivity, the
number $\xi $ associated to the (allowed) proposition $a\left( b\vee
c\right) $ can also be computed in two ways, either as $\xi \left( a\left(
b\vee c\right) \right) $ or as $\xi \left( \left( ab\right) \vee \left(
ac\right) \right) $. Therefore, using eq.(\ref{sumxi}), 
\begin{equation}
P\left( \xi \left( a\right) ,\xi \left( b\vee c\right) \right) =\text{ }\xi
\left( ab\right) +\xi \left( ac\right) ,
\end{equation}
and, using $P$ and $S$ once again, we conclude that left distributivity
leads to the following constraint 
\begin{equation}
P\left( u,v+w\right) =P\left( u,v\right) +P\left( u,w\right) ,
\label{constrps}
\end{equation}
\newline
where $\xi \left( a\right) =u$, $\xi \left( b\right) =v$, and $\xi \left(
c\right) =w$. Similarly, from propositions of the form $(a\vee b)c$ for
which right distributivity holds we obtain 
\begin{equation}
P\left( u+v,w\right) =P\left( u,w\right) +P\left( v,w\right) .
\label{constrps1}
\end{equation}

The solution of these distributivity constraints is trivial. Differentiating
eq.(\ref{constrps}) with respect to $v$ and $w$ and letting $v+w=z$ gives 
\begin{equation}
\frac{\partial ^2}{\partial z^2}\,P\left( u,z\right) =0,
\end{equation}
\newline
so that $P$ is linear in its second argument, $P(u,v)=A(u)v+B(u)$.
Substituting back into eq.(\ref{constrps}) gives $B(u)=0$. Similarly, eq.(%
\ref{constrps1}) implies that $P$ is linear in its first argument, therefore
\ \newline
\begin{equation}
P\left( u,v\right) =Cuv\quad \quad \text{or}\quad \quad \xi \left( ab\right)
=C\xi \left( a\right) \xi \left( b\right) ,
\end{equation}
The associativity constraint, eq.(\ref{constrp}) is automatically satisfied.
The constant $C$ can be absorbed into yet a new number $\psi \left( a\right)
=C\xi \left( a\right) $, so that the $and$ operation can be conveniently
represented by a simple product rule, \ 
\begin{equation}
\psi \left( ab\right) =\psi \left( a\right) \,\psi \left( b\right) ,
\label{prule}
\end{equation}
\newline
while the sum rule remains unaffected, 
\begin{equation}
\psi \left( a\vee b\right) =\psi \left( a\right) +\psi \left( b\right) .
\label{srule}
\end{equation}
\newline
Complex numbers assigned in this particularly convenient way will be called
``amplitudes''.

Let us summarize the results of this section: A quantitative representation
of the relations between setups can be obtained by assigning a complex
number $\psi(a)$ to each proposition $a$. Because of the crucial requirement
of consistency the considerable arbitrariness in the actual choice of $%
\psi(a)$ is largely illusory; it turns out that all representations are
equivalent to each other, i.e., they are obtained from each other by mere
``changes of variables.'' Although all consistent assignments are equally
correct in the sense that they serve our purpose of providing the desired
representation, some are singled out by their sheer convenience. They lead
to representations of the quantum $and$ and the $or$ operations that take
very simple forms: products and sums. This is the central result of this
paper.

\section{Wave functions and the linearity of time evolution}

Amplitudes have been introduced in the last section as the natural concept
to describe experiments quantitatively but we have not yet indicated how the
knowledge of an amplitude is to be used in predicting the outcomes of
experiments. In order to suggest, in the following section, how amplitudes
are to be interpreted, we will first explore, along conventional lines,$^{%
\text{\cite{feynman48}}}$ the properties of the amplitude $\psi (x_f,x_i)$
associated to the basic proposition $[x_f,x_i]$.

We had seen earlier, in eq. (\ref{sigmat}), how to analyze a motion from $%
x_i $ to $x_f$ in terms of motion over shorter steps from $x_i$ to $x$ and
from there to $x_f$. Now we can express this in terms of probability
amplitudes; using the sum and product rules, we get 
\begin{equation}
\psi (x_f,x_i)=\sum_{\text{all}\,\vec{x}\,\text{at}\,t}\psi
(x_f,x,x_i)=\sum_{\text{all}\,\vec{x}\,\text{at}\,t}\psi (x_f,x)\,\psi
(x,x_i),  \label{eqprop}
\end{equation}
\newline
The sums are a consequence of restricting the positions $\vec{x}$ to sites
on a discrete lattice; the generalization to a more realistic continuum
where the sums are replaced by integrals is straightforward. Equation (\ref
{eqprop}) or, more explicitly, 
\begin{equation}
\psi (\vec{x}_f,t_f;\vec{x}_i,t_i)=\sum_{\text{all}\,\vec{x}\text{ at}%
\,t}\psi (\vec{x}_f,t_f;\vec{x},t)\,\psi (\vec{x},t;\vec{x}_i,t_i),
\label{eqprop1}
\end{equation}
\newline
describes time evolution and therefore holds the key to the question of the
physical interpretation. To see this it is convenient to introduce the
notion of a state described by a wave function.

Suppose that a particle starts at $(\vec{x}_i,t_i)$ and prior to time $t$ it
undergoes various interactions the net result of which is that the amplitude
to reach the point $\vec{x}$ at time $t$ is given by $\Psi(\vec{x},t)$. Of
course, $\Psi(\vec{x},t)$ is numerically equal to $\psi(\vec{x},t;\vec{x}%
_i,t_i)$, but situations are common where we know $\Psi(\vec{x},t)$ and
either we have no interest or have lost track of what happened before $t$.
In these cases we may streamline the notation and replace $\psi(\vec{x},t;%
\vec{x}_i,t_i)$ by $\Psi(\vec{x},t)$. Using this knowledge of $\Psi(\vec{x}%
,t)$ we can calculate $\psi(x_f,x_i)$ directly. From eq. (\ref{eqprop}), 
\begin{equation}  \label{proppsi}
\psi(x_f,x_i)\equiv\Psi(\vec{x}_f,t_f)=\sum_{\text{all}\,\vec{x}\,\text{at}%
\,t}\psi(\vec{x}_f,t_f;\vec{x},t)\,\Psi(\vec{x},t)\,.
\end{equation}
\newline
The function $\Psi(\vec{x},t)$, called the wave function, represents all
those features of interactions previous to time $t$, that are relevant to
the prediction of evolution after $t$. We might, by abuse of (classical)
language, say that $\Psi$ describes the state of the particle at time $t$,
and that the effect of those interactions prior to $t$ has been to prepare
the particle in state $\Psi$.

Let's return to the description of time evolution implicit in eq. (\ref
{proppsi}). Differentiating with respect to $t_f$ and evaluating at $t_f=t$
we get 
\[
\frac{\partial \Psi (\vec{x}_f,t)}{\partial t}=\sum_{\text{all}\,\vec{x}\,%
\text{at}\,t}\frac{\partial \psi (\vec{x}_f,t^{\prime };\vec{x},t)}{\partial
t^{\prime }}\Big|_{t^{\prime }=t}\,\Psi (\vec{x},t). 
\]
\newline
The derivative on the right is a function of $\vec{x}_f$, $\vec{x}$, and of $%
t$. If we define \newline
\[
\frac{\partial \psi (\vec{x}_f,t^{\prime };\vec{x},t)}{\partial t^{\prime }}%
\Big|_{t^{\prime }=t}\equiv -\frac i\hbar \,H(\vec{x}_f,\vec{x},t), 
\]
\newline
then, 
\begin{equation}
i\hbar \frac{\partial \Psi (\vec{x}_f,t)}{\partial t}=\sum_{\text{all}\,\vec{%
x}\,\text{at}\,t}H(\vec{x}_f,\vec{x},t)\,\Psi (\vec{x},t),
\end{equation}
\newline
which is recognized as the Schr\"{o}dinger equation. We might not yet know
what $\Psi $ means, nor what should the Hamiltonian $H$ be, but we have
obtained an important result: once certain natural consistency requirements
are accepted, the time evolution of quantum states is given by a
Schr\"{o}dinger equation which is necessarily a linear equation.

The conclusion is clear: Non-linear variants of quantum mechanics that
preserve the notion of amplitudes violate natural requirements of
consistency. The question of whether it is possible to formulate non-linear
versions of quantum mechanics should not be formulated as a dynamical
question about which non-linear terms one is allowed to add to the
Schr\"odinger equation, but rather it should be phrased as a deeper
kinematical question about whether quantum mechanics should be formulated in
terms of mathematical objects other than amplitudes and wave functions.
However, whatever the nature of those mathematical objects the requirement
that they be manipulated in a consistent manner should be maintained.

\section{Physical interpretation: Born's statistical postulate.}

Having established rules for the consistent manipulation of amplitudes we
can, finally, address the question of how to use them to predict the
outcomes of experiments. The key to finding a physical interpretation for
wave functions or, equivalently, for amplitudes, is the time evolution
equation (eq. \ref{eqprop} or \ref{proppsi}). We reason as follows:

Consider first a special case. Suppose that as a result of a very special
preparation procedure (between times $t_i$ and $t$) the wave function $\Psi (%
\vec{x},t)$ at time $t$ vanishes everywhere except at a single point $\vec{x}%
_0$, 
\begin{equation}
\Psi (\vec{x},t)=A\,\delta _{\vec{x},\vec{x}_0}.
\end{equation}
\newline
Next, place a filter at time $t$ with a single hole at $\vec{x}_0$. It is
easy to see (from eq. \ref{eqprop} or \ref{proppsi}) that the presence or
absence of this filter has absolutely no influence on the subsequent
evolution of the wave function or on the amplitude to arrive at any final
destination point $x_f$. Since relations among amplitudes are meant to
reflect corresponding relations among setups, it seems natural to assume
that the presence or absence of the filter will have no effect on whether
detection at $x_f$ occurs or not. This, in turn, suggests that even in the
absence of the filter, at time $t$ the particle must have been at $\vec{x}_0$
and nowhere else. Therefore the special case where $\Psi (\vec{x},t)\propto
\delta _{\vec{x},\vec{x}_0}$ be interpreted as ``at time $t$ the particle is
located at $\vec{x}_0$.''

Essentially the same argument can be applied in a variety of other cases.
For example, if as a result of the preparation procedure $\Psi(\vec{x},t)$
vanishes at a certain point $\vec{x}^{\prime}$ then placing a filter with
holes everywhere except at $\vec{x}^{\prime}$ should have no effect on the
subsequent evolution of $\Psi$. Therefore $\Psi(\vec{x}^{\prime},t)=0$ is
interpreted as ``at time $t$ the particle is not at $\vec{x}^{\prime}$.''

This leads naturally to the following general interpretative postulate:
Consider a filter the action of which is to block out those components of
the wave function characterized by a certain feature (what that feature is
should be obvious to whoever built the filter). Suppose that at time $t$ the
system is in a state of wave function $\Psi\left(t\right)$. If the
introduction or removal of the filter at $t$ has no effect on the future
evolution of the wave function then the rule of interpretation is that the
system in state $\Psi\left(t\right)$ does not exhibit the feature in
question. The rule applies to amplitudes in general.

The power of this rule becomes apparent when applied to a particle with$\,$a
generic wave function$\,$ 
\begin{equation}
\Psi (\vec{x},t)=\sum_iA_i\,\delta _{\vec{x},\vec{x}_i},  \label{psi}
\end{equation}
\newline
where the number and location of the $\vec{x}_i$s is arbitrary. We want to
predict the outcome of an experiment in which a detector is placed at a
certain $\vec{x}_k$. If $\vec{x}_k$ differs from all of the $\vec{x}_i$s in
the sum in eq. (\ref{psi}) the interpretative rule directly implies that the
particle will definitely not be detected.

The interesting problem arises when $\vec{x}_k$ coincides with one of the $%
\vec{x}_i$s. In this case, as expected, one cannot predict the actual
outcome of the experiment. What is predictable, and quite precisely in fact,
is the probability of various outcomes. At this point we will make an
important assumption about the wave function: we will assume that it can be
normalized. For convenience we will from now on assume that $\Psi $ in eq. (%
\ref{psi}) has been appropriately normalized, 
\begin{equation}
\left\| \Psi \right\| ^2=\left( \Psi ,\Psi \right) \equiv \sum_{\vec{x}%
}\left| \Psi (\vec{x},t)\right| ^2=\sum_i|A_i|^2=1.
\end{equation}
Next we show that the probability of detection at $\vec{x}_k$ is $|A_k|^2$.
Thus Born's statistical interpretation is actually a theorem: it follows
from a simpler interpretative rule that only refers to situations of
absolute certainty. This remarkable fact was discovered long ago by
Finkelstein,$^{\text{\cite{finkelstein63}}}$ by Hartle$^{\text{\cite
{hartle68}}}$ and by Graham.$^{\text{\cite{graham73}}}$ The proof below is
particularly suited to the approach to quantum theory being developed in
this paper.

Consider an ensemble of $N$ identically prepared, independent replicas of
our particle; later we will take $N\rightarrow \infty $. In the next section
we will show that the wave function for this $N$-particle system is the
product 
\begin{equation}
\Psi _N(\vec{x}_1,\ldots ,\vec{x}_N,t)=\prod_{\alpha =1}^N\Psi (\vec{x}%
_\alpha ,t)=\sum_{i_1\ldots i_N}A_{i_1}\ldots A_{i_N}\,\delta _{\vec{x}_1,%
\vec{x}_{i_1}}\ldots \delta _{\vec{x}_N,\vec{x}_{i_N}}.
\end{equation}
Suppose that in the $N$-particle configuration space we place a special
filter, denoted by $P_n^k$, the action of which is to block all components
of $\Psi _N$ except those where exactly $n$ of the $N$ replicas are at $\vec{%
x}_k$. The wave function right after this filter is 
\begin{equation}
P_n^k\Psi _N=\sum_{i_1\ldots i_N}\delta _{n,n_k}\,A_{i_1}\ldots
A_{i_N}\,\delta _{\vec{x}_1,\vec{x}_{i_1}}\ldots \delta _{\vec{x}_N,\vec{x}%
_{i_N}},
\end{equation}
where 
\begin{equation}
n_k=\sum_{\alpha =1}^N\delta _{k,i_\alpha }.
\end{equation}
\newline
Actually this filter is too strict, it selects a single sharply-defined
fraction $f=n/N$. What we need is a more lenient filter (presumably built by
opening additional ``holes'' in $P_n^k$) that allows passage of all
fractions in a range from $f-\epsilon $ to $f+\epsilon $. The action of this
filter is described by 
\begin{equation}
P_{f,\epsilon }^k\Psi _N=\sum_{i_1\ldots i_N}\left( \sum_{n=(f-\epsilon
)N}^{(f+\epsilon )N}\delta _{n,n_k}\right) \,A_{i_1}\ldots A_{i_N}\,\delta _{%
\vec{x}_1,\vec{x}_{i_1}}\ldots \delta _{\vec{x}_N,\vec{x}_{i_N}}\text{ .}
\end{equation}

We are now ready to apply our interpretative rule: If, as $N\rightarrow
\infty $, the presence of the filter $P_{f,\epsilon }^k$ is found to have no
influence whatsoever on the future evolution of the wave function $\Psi _N$
we will interpret $\Psi _N$ as representing a state for which the fractions
of replicas at $\vec{x}_k$ must lie in the range from $f-\epsilon $ to $%
f+\epsilon $. To show that this is actually the case we must show that as $%
N\rightarrow \infty $ the wave function right after the filter, $%
P_{f,\epsilon }^k\Psi _N$, becomes more and more similar to the wave
function right before the filter, $\Psi _N$. To this end we calculate the
norm \newline
\begin{equation}
\left\| P_{f,\epsilon }^k\Psi _N-\Psi _N\right\| ^2=\sum_{\vec{x}_1\ldots 
\vec{x}_N}\left| P_{f,\epsilon }^k\Psi _N-\Psi _N\right| ^2.
\end{equation}
\newline
Since filters act as projectors, $PP=P$, we get \ 
\begin{equation}
\left\| P_{f,\epsilon }^k\Psi _N-\Psi _N\right\| ^2=1-\left( \Psi
_N,P_{f,\epsilon }^k\Psi _N\right) .
\end{equation}
\newline
The calculation of the scalar product is straightforward, 
\begin{equation}
\left( \Psi _N,P_{f,\epsilon }^k\Psi _N\right) =\sum_{\vec{x}_1\ldots \vec{x}%
_N}\Psi _N^{*}\,P_{f,\epsilon }^k\Psi _N=\sum_{n=(f-\epsilon
)N}^{(f+\epsilon )N}\left( \sum_{i_1\ldots i_N}\delta _{n,n_k}\,\left|
A_{i_1}\right| ^2\ldots \left| A_{i_N}\right| ^2\right) .
\end{equation}
\newline
The sum over $i_1$,$\ldots $,$\,i_N$ is done as follows: Suppose we satisfy
the Kronecker $\delta _{n,n_k}$ constraint by choosing $n$ of the $N$
indices $i_1$,$\ldots $,$\,i_N$ and setting them to the value $k$. Since the
individual $\Psi $s are normalized each sum over the remaining $N-n$ indices
gives 
\begin{equation}
\sum_{i\neq k}|A_i|^2=1-|A_k|^2\text{ .}
\end{equation}
\newline
But there are $\binom Nn$ ways to choose which $n$ indices are set equal to $%
k$, therefore 
\begin{equation}
\left( \Psi _N,P_{f,\epsilon }^k\Psi _N\right) =\,\,\,\sum_{n=(f-\epsilon
)N}^{(f+\epsilon )N}\binom Nn\left( |A_k|^2\right) ^n\left( 1-|A_k|^2\right)
^{N-n}.
\end{equation}
\newline
For large $N$ this binomial sum tends to the integral of a Gaussian, 
\begin{equation}
\left( \Psi _N,P_{f,\epsilon }^k\Psi _N\right) =\,\,\,\int_{f-\epsilon
}^{f+\epsilon }\frac 1{\sqrt{2\pi \sigma _N^2}}\,exp\left( \frac{\left(
f^{\prime }-\overline{f}\right) ^2}{2\sigma _N^2}\right) df^{\prime }.
\end{equation}
\newline
with mean $\overline{f}=|A_k|^2$ and variance $\sigma _N^2=\overline{f}(1-%
\overline{f})/N$. In the limit $N\rightarrow \infty $ this is more concisely
written as a $\delta $ function, therefore \ 
\begin{equation}
\stackunder{N\rightarrow \infty }{lim}\,\left\| P_{f,\epsilon }^k\Psi
_N-\Psi _N\right\| ^2\equiv 1-\,\,\,\int_{f-\epsilon }^{f+\epsilon }\delta
\left( f^{\prime }-|A_k|^2\right) df^{\prime }.
\end{equation}
The interpretation is clear: as $N\rightarrow \infty $ the filter $%
P_{f,\epsilon }^k$ will have no effect on the wave function $\Psi _N$
provided $f$ lies in a range $2\epsilon $ about $|A_k|^2$, and according to
our interpretative rule $\Psi _N$ cannot contain any fractions outside this
range. Choosing stricter filters with $\epsilon \rightarrow 0$ we conclude
that as $N\rightarrow \infty $, $\Psi _N$ is a state for which the fraction
of replicas at $\vec{x}_k$ is exactly $|A_k|^2$.

Returning to the original single particle system, we see that we cannot
predict whether a detection at $\vec{x}_k$ will occur or not. In fact, we
have a very definite prediction of indeterminism: for large $N$ detection
will certainly occur for a fraction $|A_k|^2$, and equally significant,
detection will certainly not occur for a fraction $1-|A_k|^2$. Once the
assumption is made that the relations among possible experimental setups are
quantitatively represented by consistently assigned amplitudes, the general
interpretative rule implies indeterminism. The best one can do is assign a
probability $p$ to this detection. Given the information that for a large
number of identically prepared systems the fraction of successful detections
is $|A_k|^2$ the only assignment consistent with the law of large numbers is
the value 
\begin{equation}
p\text{ }=|A_k|^2\,.
\end{equation}
\newline
To complete our proof of Born's postulate we must prove that the wave
function for $N$ independent particles is the product of the wave functions
for each one of the particles. This is the topic of the next section.

\section{Several independent particles}

We want to show that the wave function of a system $\alpha \beta $ composed
of two independent particles $\alpha $ and $\beta $ is the product of the
wave functions for each particle, $\Psi _{\alpha \beta }=\Psi _{\alpha
{}}\Psi _{\beta {}}$. In the spirit of the previous sections the first step
must be that of defining the statements about composite systems in terms of
the experimental setups designed to test them and of providing a
representation in terms of amplitudes of the relations among those setups.

Our system is composed of two independent particles. The notion of
independence imposes highly non-trivial constraints. Suppose that the
allowed propositions about particle $\alpha$ by itself and the corresponding
setups designed to test them are denoted by $a$, as in eq. (1), and
similarly, that the allowed propositions and setups about particle $\beta$
are denoted by $b$. Then the first condition implied by independence is that
the statements $c$ about the composite $\alpha\beta$ are restricted those
that can be tested by composite setups that separately test $a$ about $%
\alpha $ and $b$ about $\beta$. Thus the setups allowed for $\alpha\beta$
are of the form $c=\{a;b\}$.

The various ways in which the composite setups $c$ can be combined can be
derived from the various ways in which the $a$s and the $b$s can be combined
among themselves. Thus, if $c_1=\{a_1;b_1\}$ and $c_2=\{a_2;b_2\}$ and if $%
a_1\vee a_2$ is allowed and $b_1=b_2$ then we define $or$ by $c_1\vee
c_2\equiv\{a_1\vee a_2,b_1\}$. On the other hand if it is $b_1\vee b_2$ that
is allowed and $a_1=a_2$ then $c_1\vee c_2\equiv\{a_1,b_1\vee b_2\}$. In the
general case $c_1\vee c_2$ will be an allowed setup only if $c_1$ and $c_2$
differ in one and only one filter; if the $b$ setups differ then the $a$s
must be identical and vice versa. Similarly, if both $a_1a_2$ and $b_1b_2$
are allowed then we define $c_1c_2\equiv\{a_1a_2,b_1b_2\}$. Commutativity,
associativity and distributivity for the $and$ and $or$ relations among the $%
c$ setups follow from the corresponding properties for the $a$ and $b$
setups. The argument of section 3 can now be repeated: the relations among
different composite setups can be conveniently represented quantitatively by
assigning a complex amplitude $\psi(c)$ to each setup $c$ in such a way that
the sum and product rules hold.

The second crucial condition implicit in the notion of independence, one
that goes beyond the mere capability of independently placing filters in the
paths of $\alpha $ and $\beta $, is the requirement that changing the
filters acting on $\alpha $, i.e., changing $a$ to $a^{\prime }$ shall have
no influence whatsoever on the outcome of $b$, and vice versa. Since
relations among amplitudes are meant to reflect corresponding relations
among setups, and the physically relevant information about setups $a$ and $%
b $ is contained in $\psi (a)$ and $\psi (b)$ this second condition can be
quantitatively expressed by the requirement that the physically relevant
information about setup $c$, expressed by $\psi (c)$ be some function of $%
\psi (a)$ and $\psi (b)$ and nothing else. Thus, 
\begin{equation}
\psi \left( c\right) =F\left( \psi \left( a\right) ,\psi \left( b\right)
\right) .
\end{equation}
\newline
This is what we mean by independence.$^{\text{\cite{entangle}}}$

The function $F$ is determined from the fact that not only $\psi (a)$ and $%
\psi (b)$ but also $\psi (c)$ must satisfy the sum and product rules. For
example, if $c_1=\{a_1;b_1\}$ and $c_2=\{a_2;b_2\}$ and if $a_1\vee a_2$ is
allowed and $b_1=b_2$ then $\psi (c_1\vee c_2)=\psi (c_1)+\psi (c_2)$
implies \ 
\begin{equation}
F\left( \psi \left( a_1\right) +\psi \left( a_2\right) ,\psi \left(
b_1\right) \right) =F\left( \psi \left( a_1\right) ,\psi \left( b_1\right)
\right) +F\left( \psi \left( a_2\right) ,\psi \left( b_1\right) \right) ,
\end{equation}
\newline
so that 
\begin{equation}
F\left( u+v,w\right) =F\left( u,w\right) +F\left( v,w\right) .
\label{findep}
\end{equation}
\newline
Similarly, if $b_1\vee b_2$ is allowed and $a_1=a_2$ then 
\begin{equation}
F\left( u,v+w\right) =F\left( u,v\right) +F\left( u,w\right) \,.
\label{findep1}
\end{equation}
\newline
Finally, if $c_1c_2\equiv \{a_1a_2,b_1b_2\}$ is allowed, \ from the product
rule $\psi (c_1c_2)=\psi (c_1)\psi (c_2)$ we get 
\begin{equation}
F\left( uv,wz\right) =F\left( u,w\right) F\left( v,z\right) .
\label{findep2}
\end{equation}
\newline
Equations (\ref{findep}) and (\ref{findep1}) are formally identical with
equations (\ref{constrps}) and (\ref{constrps1}). Therefore $F(u,v)=Cuv$.
Substituting into (\ref{findep2}) we get $C=1$, $F(u,v)=uv$.

Therefore, if particles $\alpha $ and $\beta $ are independent, the
amplitude associated to $c=\{a;b\}$ is the product of the amplitudes
associated to $a$ and to $b$, 
\begin{equation}
\psi (c)=\psi (a)\psi (b).
\end{equation}
\ \newline
The generalization to $N$ independent particles is straightforward.

\section{Conclusions and some comments}

Let us summarize our main conclusions: After having identified a restricted
set of allowed propositions in terms of the experimental setups designed to
test them we introduce amplitudes as the essentially unique tool to carry
out consistent calculations. The rules of manipulation are necessarily such
that time evolution is described by a linear Schr\"odinger equation and that
the Born probability interpretation holds.

It might seem surprising that a substantial amount of quantum mechanics has
been reproduced without any mention of commutation relations among
incompatible observables and of the corresponding uncertainty relations; in
fact, we have only discussed the measurement of position. However,
observables other than position are useful concepts in that they facilitate
the description of experiments and the manipulation of information. Although
they have not been central in this formulation of quantum theory it may be
worthwhile to remark on how they arise. In general, they originate from the
idea that by placing various filters, diffraction gratings, magnetic fields,
etc., prior to the final position detection at $x_f$ one can effectively
build a more complex detector. This raises two questions; the first is what
does such a complex device actually measure. The answer$^{\text{\cite
{feynman48}}}$ is that what is measured is the extent to which the actual
wave function $\Psi $ resembles another wave function $\Phi $ which is a
characteristic of the detector. Should the resemblance be complete the
particle would be detected at $x_f$ with absolute certainty. The second
question is what properties are we actually interested in measuring.
Typically, interesting measurements will result in information useful for
prediction in other experiments, and foremost among these is the measurement
of properties that have some lasting value, i.e., conserved quantities. A
related question is that of deciding how the Hamiltonian should be chosen;
this choice, like that of most other observables, is dictated by symmetries$%
^{\text{\cite{jauch68},\cite{weinberg95}}}$ and will not be further pursued
here.

For clarity we have focused our attention on the special case of a single
particle moving in a discrete lattice. But it should be clear that the
argument can be generalized to considerably more complicated configuration
spaces. The crucial feature is to identify the relevant propositions or
equivalently, the idealized experimental setups designed to test them, and
verify that the appropriate rules of associativity and distributivity hold.
It is of some interest that in order to implement the associativity
constraint one requires a configuration space which consists at least of
three values. Remarkably, a similar restriction to spaces of three
dimensions or more appears also in the work of Gleason.$^{\text{\cite
{gleason57}}}$ This does not, of course, represent a problem: two-valued
configuration spaces are unphysical. For example, realistic spin-$1/2$
systems also have translational and \ a variety of other degrees of freedom.

It is possible that the use of more complex detectors (i.e., other
observables) will permit extending the set of allowed propositions. Perhaps
this would bridge the gap between the present formulation and the quantum
logic approach.$^{\text{\cite{jauch68},\cite{hooker79},\cite{finkelstein63}}%
} $ However, whether such an extension is advantageous is not at all clear.
It would surely spoil the distributivity property (of $and$ and $or$) which
has played such a crucial role here.

Many are the questions left open. Most, like the application of quantum
mechanics to the detectors themselves, as well as to other macroscopic
objects, the nature of the classical limit, and other issues associated with
decoherence through interaction with the environment, are common to all
approaches to quantum theory. But some questions seem more urgent in this
formulation. A particularly glaring one is: why complex numbers? Perhaps
other mathematical objects with the appropriate associative and distributive
algebras (e.g., quaternions, multivectors, etc.)$^{\text{\cite{finkelstein62}%
}}$ could be used to obtain representations of the $and$ and $or$ operations.

We conclude with a couple of brief comments. The first concerns the
similarity between quantum motion and Brownian motion, or between the
Schr\"odinger equation and the diffusion equation. There is a natural
reticence to dismiss it as a mere coincidence, and it has been suggested
that perhaps there is some underlying stochastic physical agent responsible
for the peculiar features of quantum motion. Our results suggest that such a
physical agent need not exist, that the similarities between quantum and
Brownian theories arise from formalisms that are strongly constrained by
similar logical requirements of consistency which force one to manipulate
amplitudes in one case, and probabilities in the other, using similar sum
and product rules.

The second comment addresses another aspect of the robustness of quantum
theory, its universality. Quantum mechanics applies to a wide variety of
systems over a broad range of energy and distance scales. Were new exotic
objects (say, new excitations in condensed matter, or new particles, or
strings) to be discovered, could we expect them to be described by quantum
mechanics? Classical mechanics fails at atomic scales; how short a distance
can we go and still expect quantum mechanics to hold? Or, in other words:
What are the accepted features of today's physics that could reasonably be
expected to hold in the future physics of objects that are yet to be
discovered, or of energy and distance scales that are yet to be explored? It
seems natural to assume that any list of such features should prominently
include those derivable from mere consistency requirements; linearity and
indeterminism are likely to be among them. \ 

\appendix 

\section{Solution of the consistency equations}

Our approach to solving the associativity equation (\ref{constrs}), 
\begin{equation}
S\left( S\left( u,v\right) ,w\right) =S\left( u,S\left( v,w\right) \right) ,
\label{a1}
\end{equation}
\newline
is essentially that due to Cox.$^{\text{\cite{cox46}}}$ A minor difference
is that we deal with complex rather than real variables. Let $r=S\left(
u,v\right) $, $s=S\left( v,w\right) $, $S_1(u,v)=\partial S\left( u,v\right)
/\partial u$, and $S_2(u,v)=\partial S\left( u,v\right) /\partial v$. Then
eq.(\ref{a1}) and its derivatives with respect to $u$ and $v$ are 
\begin{equation}
S\left( r,w\right) =S\left( u,s\right) ,
\end{equation}
\begin{equation}
S_1(r,w)S_1(u,v)=S_1(u,s),
\end{equation}
and 
\begin{equation}
S_1(r,w)S_2(u,v)=S_2(u,s)S_1(v,w).
\end{equation}
\newline
Eliminating $S_1(r,w)$ from these last two equations we get 
\begin{equation}
G(u,v)=G(u,s)S_1(v,w).  \label{g2}
\end{equation}
\newline
where 
\begin{equation}
G(u,v)=\frac{S_2(u,v)}{S_1(u,v)}.  \label{g3}
\end{equation}
\newline
Multiplying eq.(\ref{g2}) by $G(v,w)$ we get 
\begin{equation}
G(u,s)G(v,w)=G(u,s)S_2(v,w)\text{ }  \label{g4}
\end{equation}
\newline
Differentiating the right hand side of eq.(\ref{g4}) with respect to $v$ and
comparing with the derivative of eq.(\ref{g2}) with respect to $w$, we have 
\begin{equation}
\frac \partial {\partial v}\left( G\left( u,s\right) S_2\left( v,w\right)
\right) =\frac \partial {\partial w}\left( G\left( u,s\right) S_1\left(
v,w\right) \right) =\frac \partial {\partial w}\left( G\left( u,v\right)
\right) =0.
\end{equation}
\newline
Therefore 
\begin{equation}
\frac \partial {\partial v}\left( G\left( u,v\right) G\left( v,w\right)
\right) =0,
\end{equation}
or, 
\begin{equation}
\frac 1{G\left( u,v\right) }\frac{\partial G\left( u,v\right) }{\partial v}%
=-\frac 1{G\left( v,w\right) }\frac{\partial G\left( v,w\right) }{\partial v}%
\equiv h\left( v\right) .  \label{a5}
\end{equation}
\newline
Integrating, we get 
\begin{equation}
G(u,v)=G(u,0)\,exp\,\int_0^vh(v^{\prime })dv^{\prime },
\end{equation}
\newline
and also 
\begin{equation}
G\left( v,w\right) =G\left( 0,w\right) \,exp\,-\int_0^vh(v^{\prime
})dv^{\prime },
\end{equation}
\newline
so that 
\begin{equation}
G\left( u,v\right) =c\,\frac{H\left( u\right) }{H\left( v\right) },
\end{equation}
where $c=G(0,0)$ is a constant and \newline
\begin{equation}
H(u)=exp\,-\int_0^uh(u^{\prime })du^{\prime }.  \label{a6}
\end{equation}
\newline
On substituting back into eqs.(\ref{g2}) and (\ref{g4}) we get 
\begin{equation}
S_1(v,w)=\,\frac{H(s)}{H(v)}\quad \quad \text{and}\quad \quad S_2(v,w)=c\,%
\frac{H(s)}{H(w)}.  \label{a7}
\end{equation}
\newline
But $s=S(v,w)$, so substituting (\ref{a7}) into $ds=S_1(v,w)dv+S_2(v,w)dw$
we get \ \ 
\begin{equation}
\frac{ds}{H(s)}=\frac{dv}{H(v)}+c\frac{dw}{H(w)}.
\end{equation}
This is easily integrated. Let 
\begin{equation}
\xi \left( u\right) =\xi \left( 0\right) +\int_0^u\frac{du^{\prime }}{%
H(u^{\prime })},
\end{equation}
so that $du/H(u)=d\xi (u)$. Then 
\begin{equation}
\xi \left( S\left( v,w\right) \right) =\xi \left( v\right) +c\xi \left(
w\right) ,
\end{equation}
where a constant of integration has been absorbed into $\xi \left( 0\right) $%
. Substituting this function $\xi $ back into eq.(\ref{a1}) we obtain $c=1$.
This completes our derivation of eq.(\ref{xi1}).

\noindent \textbf{Acknowledgements}

I am indebted to C. Rodriguez, P. Zambianchi, A. Inomata, and J. Kimball for
valuable discussions and many insightful remarks.


\begin{thebibliography}{99}
\bibitem{entangle}  The two conditions implicit in the notion of
independence address the two ways in which the wave function of two systems
can become entangled: either during the preparation or through interaction.
The challenge here is to implement these conditions without using tools
which are available in the usual formulations (ideas of entanglement,
interaction potentials, etc.) but are not yet available to us.
\end{thebibliography}
\end{document}